\documentclass[prl,print,showpacs,superscriptaddress,twocolumn]{revtex4}
\usepackage[dvips]{graphics,color}
\usepackage{amsfonts}
\usepackage{amsmath}
\usepackage{amssymb}
\usepackage{graphicx}

\begin{document}

\title{Finite temperature Dicke phase transition of a Bose-Einstein
condensate in an optical cavity}
\author{Yuanwei Zhang}
\thanks{These authors contributed equally to this work}
\affiliation{Institute of Theoretical Physics, Shanxi University, Taiyuan 030006, P. R.
China}
\author{Jinling Lian}
\thanks{These authors contributed equally to this work}
\affiliation{Institute of Theoretical Physics, Shanxi University, Taiyuan 030006, P. R.
China}
\author{J. -Q. Liang}
\affiliation{Institute of Theoretical Physics, Shanxi University, Taiyuan 030006, P. R.
China}
\author{Gang Chen}
\thanks{Corresponding author, chengang971@163.com}
\affiliation{State Key Laboratory of Quantum Optics and Quantum Optics Devices, College
of Physics and Electronic Engineering, Shanxi University, Taiyuan 030006, P.
R. China}
\author{Chuanwei Zhang}
\affiliation{Department of Physics and Astronomy, Washington State University, Pullman,
Washington, 99164 USA}
\author{Suotang Jia}
\affiliation{State Key Laboratory of Quantum Optics and Quantum Optics Devices, College
of Physics and Electronic Engineering, Shanxi University, Taiyuan 030006, P.
R. China}

\begin{abstract}
Dicke model predicts a quantum phase transition from normal to superradiant
phases for a two-level atomic ensemble coupled with an optical cavity at
zero temperature. In a recent pioneer experiment [Nature \textbf{464}, 1301
(2010)], such a phase transition has been observed using a Bose-Einstein
condensate (BEC) in an optical cavity. Compared with the original Dicke
model, the experimental system features finite temperature and strong
atom-photon nonlinear interaction in BEC. In this Letter, we develop a
finite temperature theory for the Dicke phase transition of a BEC in an
optical cavity, taking into account the atom-photon nonlinear interaction.
In addition to explaining the experimentally observed transition from normal
to superradiant phases at finite-temperature, we point it out that a new
phase, the coexistence of normal and superradient phases, was also observed
in the experiment. We show rich finite temperature phase diagrams existing
in the experimental system by tuning various experimental parameters. We
find that the specific heat of the BEC can serve as a powerful tool for
probing various phases.
\end{abstract}

\maketitle

The Dicke model, a textbook paradigm in quantum optics, describes a
two-level atomic ensemble coupled with an optical cavity. It was first
introduced to illustrate the importance of collective and coherent
excitations for atoms induced by a single photon mode \cite{Dicke}. With
increasing atom-cavity coupling, the Dicke model predicts a quantum phase
transition from a normal phase (NP) to a superradiant phase (SP), where both
the atomic ensemble and photon acquire macroscopically collective
excitations \cite{Hepp,Wang,Hioes}. However, due to the \textbf{`}no-go
theorem\textbf{' }originating from the Thomas-Reiche-Kuhn sum rule for the
oscillator strength, this phase transition cannot be realized in typical
cavity quantum electrodynamics \cite{PN,OV}. The experimental breakthrough
for observing the quantum phase transition only occurs recently using the
momentum eigenstates of a Bose-Einstein condensate (BEC) coupled with an
optical cavity \cite{Baumann1,Baumann2}. In this pioneer experiment, there
exists an nonlinear atom-photon interaction \cite{Nagy}, which, induced by
the optical lattice potential and greatly enhanced by the large atom number $%
N$, can reach the same order of the effective cavity frequency and even go
beyond. In this strong interaction regime, rich dynamical properties \cite%
{Keeling,Bhaseen} as well as new quantum phase transitions \cite{Chen2} have
been predicted at zero temperature.

Although such ideal quantum phase transitions should occur at absolute zero
temperature \cite{Sachdev} in principle, realistic experiments used to
observe the quantum phase transition must be performed at finite
temperature. For instance, the typical temperature of the BEC in \cite%
{Baumann1,Baumann2} for observing the Dicke phase transition is $\sim 50$
nK. At finite temperature, thermal fluctuations may induce new exotic
phenomena beyond the prediction of the zero temperature theory \cite{BEC}. A
well-known example is the lost of long range superfluid order in low
dimensions (two or one) at any temperature \cite{PCH}, where the superfluid
physics is characterized by the Berezinskii-Kosterlitz-Thouless transition
\cite{VLB,KT}. Therefore it is crucially important to investigate the Dicke
phase transition at finite temperature to fully understand the realistic
experiment.

In this Letter, we present a finite-temperature field theory for the Dicke
phase transition in a BEC coupled with an optical cavity to understand the
recent breakthrough experiment in this system. Our main findings are the
following:

(I) The experimentally observed phase diagram for the normal-superradiant
phase transition, (\emph{i.e.}, Fig. (5) in Ref. \cite{Baumann1}) can be
well understood in our finite-temperature theory. More interestingly, we
show that a new phase, called CE$_{\text{SP}}$, can be identified in the
experimental phase diagram in \cite{Baumann1}. In the CE$_{\text{SP}}$
phase, NP and SP coexist, but the SP is stable and the NP is metastable.

(II) By varying various physical parameters (temperature, the coupling
strength, the nonlinear interaction, etc.), we show there exist rich
finite-temperature phases in the experimental system, including NP, SP, CE$_{%
\text{SP}}$, CE$_{\text{NP}}$ (coexistence of stable NP and metastable SP),
as well as dynamical unstable phases (US). In certain parameter region,
there is a four-phase coexistence point.

(III) We show, both analytically and numerically, that the specific heat of
the BEC increases rapidly in the SP/CE$_{\text{SP}}$ (exponential increase
at low temperature) with increasing temperature, and has a large jump at the
critical transition temperature where the system becomes the NP/CE$_{\text{NP%
}}$. Therefore the specific heat of the BEC may serve as a powerful tool for
detecting different phases in the temperature-driven Dicke phase transition.

\begin{figure}[tp]
\includegraphics[width=8cm]{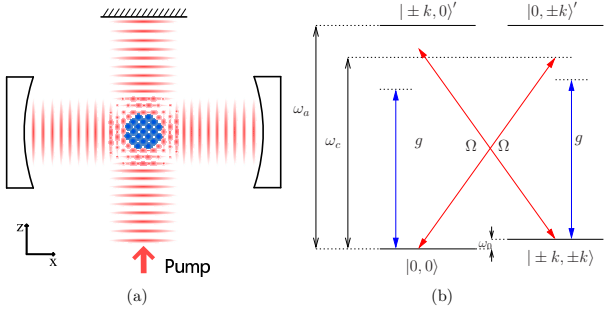}\newline
\caption{(Color online) (a) A BEC with the $^{87}$Rb atoms interacts
strongly with a high-finesse optical cavity. (b) A four-level model is
considered by introducing the zero momentum state $\left\vert
0,0\right\rangle (=\left\vert p_{x},p_{z}\right\rangle $ with $p_{x}$ and $%
p_{z}$ being the momenta in the $x$ and $z$ directions, hereafter$)$, the
excited states $\left\vert \pm k,0\right\rangle $ and $\left\vert 0,\pm
k\right\rangle ,$ and the symmetric superposition of states $\left\vert \pm
k,\pm k\right\rangle $. In the dispersive limit, the excited-state levels
can be eliminated adiabatically, and an effective two-level system with the
zero momentum state and the symmetric superposition of states is thus
formed. }
\label{fig1}
\end{figure}

Figure \ref{fig1} shows the experimental setup in which all atoms in a $^{87}
$Rb BEC interact identically with a single-mode photon induced by a
high-finesse optical cavity \cite{Baumann1}. In the experimental scheme, the
atom momenta are used to define the spin states $\left\vert \uparrow
\right\rangle \equiv \left\vert \pm k,\pm k\right\rangle $, $\left\vert
\downarrow \right\rangle \equiv \left\vert 0,0\right\rangle $ with the
corresponding SU(2) collective spin operators $S_{+}=S_{-}^{\dagger
}=\sum_{i}\left\vert \uparrow \right\rangle _{ii}\left\langle \downarrow
\right\vert $ and $S_{z}=\sum_{i}(\left\vert \uparrow \right\rangle
_{ii}\left\langle \uparrow \right\vert -\left\vert \downarrow \right\rangle
_{ii}\left\langle \downarrow \right\vert )$, as shown in Fig. \ref{fig1}(b).
As a result, not only the `no-go theorem' is overcome, but also the
superradiant-normal phase transition condition can be satisfied \cite%
{Emary,Dimer,Chen1}, since the smaller energy scale of the effective two
levels can be achieved. Under the new spin basis, the dynamics of the
atom-cavity system are governed by the Hamiltonian \cite{Baumann1}
\begin{equation}
H=(\omega +\frac{U}{2N}S_{z})\psi ^{\dagger }\psi +\frac{\omega _{0}}{2}%
S_{z}+\frac{g}{\sqrt{N}}(\psi +\psi ^{\dagger })(S_{+}+S_{-}).
\label{Total H}
\end{equation}%
Henceforth we set $\hbar =1$. $\psi ^{\dagger }$ denotes the photon creation
operator. The effective cavity frequency $\omega =-\Delta _{c}+NU_{0}(1+\Xi
)/2$ with $\Delta _{c}=\omega _{p}-\omega _{c}$, where $\omega _{c}$ is the
cavity frequency, $\omega _{p}$ is the pump laser frequency, $\Xi =3/4$, and
$U_{0}=g_{0}^{2}/(\omega _{p}-\omega _{a})$ with $g_{0}$ being the coupling
strength between a single atom and the photon and $\omega _{a}$ being the
atomic transition frequency. The nonlinear atom-photon interaction $U=N\Xi
U_{0}$, which is greatly enhanced by the large atom number $N$, and can
reach the same order as the effective cavity frequency $\omega $ or even
beyond. Therefore a strong nonlinear atom-photon interaction regime is
accessible in experiments. The effective atomic frequency $\omega
_{0}=2\omega _{r}$ with the atomic recoil energy $\omega _{r}$ $=k^{2}/2m$.
The collective coupling strength $g=g_{0}\Omega \sqrt{N}/2(\omega
_{p}-\omega _{a})$, and $\Omega $ is the maximum pump Rabi frequency\textbf{.%
}

\begin{figure}[b]
\includegraphics[width=8cm]{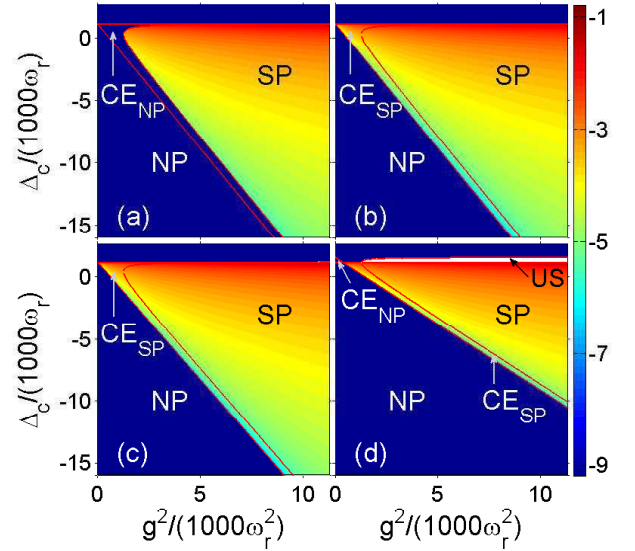}\newline
\caption{(Color online) The scaled mean-photon number as the functions of
the detuning $\Delta _{c}$ and the square of the collective coupling
strength $g$ for the temperature $T=0.00$ nk (a), $T=0.13$ nk (b) $T=50.00$
nk (c), and $T=150.00$ nk (d). The nonlinear interaction $U=1.70\times 10^{3}%
\protect\omega _{r}$ and the recoil frequency $\protect\omega _{r}=2\protect%
\pi \times 3.71$ kHz. 1 nk $\leftrightarrow 130.00$ Hz.}
\label{fig2}
\end{figure}

The finite-temperature properties of the BEC-cavity system can be obtained
by calculating the partition function of the system through the
imaginary-time functional path-integral approach. In this procedure, we
rewrite the collective spin operators in the Hamiltonian (\ref{Total H}) as $%
S_{z}=\sum_{i=1}^{N}(\alpha _{i}^{\dag }\alpha _{i}-\gamma _{i}^{\dag
}\gamma _{i})$ and $S_{+}=\sum_{i=1}^{N}\alpha _{i}^{\dag }\gamma _{i}$
using Fermi operators $\alpha _{i}^{\dag }(\alpha _{i})$ and $\gamma
_{i}^{\dag }(\gamma _{i})$, which obey the anticommutation relation $%
\{\alpha _{i}^{\dag },\alpha _{j}\}=\{\gamma _{i}^{\dag },\gamma
_{j}\}=\delta _{ij}$. After a straightforward calculation, the effective
partition function can be written as \cite{Popov}%
\begin{equation}
\frac{Z}{Z_{0}}=\frac{\int [d\eta ]\exp \left( -S\right) }{\int [d\eta ]\exp
(-S_{f})},  \label{PT}
\end{equation}%
where $[d\eta ]$ is the functional measure. In Eq. (\ref{PT}), the free
action of the bosonic field $S_{f}=\int_{0}^{\beta }d\tau \sum_{i=1}^{N}%
\left[ \alpha _{i}^{\ast }\left( \tau \right) \partial _{\tau }\alpha
_{i}\left( \tau \right) +\gamma _{i}^{\ast }\left( \tau \right) \partial
_{\tau }\gamma _{i}\left( \tau \right) \right] +\int_{0}^{\beta }d\tau \psi
^{\ast }\left( \tau \right) \left( \partial _{\tau }+\omega \right) \psi
\left( \tau \right) $, where $\tau =it$, $\partial _{\tau }=$ $\partial
/\partial \tau $, $\beta =1/\left( k_{B}T\right) $ with $k_{B}$ being the
Boltzmann constant and $T$ being the system's temperature. The total action $%
S$ is expressed, in the basis of $\Phi _{i}\left( \tau \right) =\left[
\gamma _{i}\left( \tau \right) ,\alpha _{i}\left( \tau \right) \right] ^{T}$%
, as
\begin{equation}
S=S_{0}\left( \psi ,\psi ^{\ast }\right) +\int_{0}^{\beta }d\tau \Phi
_{i}^{\dag }\left( \tau \right) G\left( \psi ,\psi ^{\ast }\right) \Phi
_{i}\left( \tau \right) ,  \label{EA}
\end{equation}%
where $S_{0}\left( \psi ,\psi ^{\ast }\right) =\int_{0}^{\beta }d\tau \psi
^{\ast }\left( \tau \right) \partial _{\tau }\psi \left( \tau \right) $ and%
\begin{equation}
G=\left(
\begin{array}{cc}
\partial _{\tau }+\frac{\omega _{0}}{2}+\frac{U}{2N}\psi ^{\ast }\psi &
\frac{g}{\sqrt{N}}\left( \psi ^{\ast }+\psi \right) \\
\frac{g}{\sqrt{N}}\left( \psi ^{\ast }+\psi \right) & \partial _{\tau }-%
\frac{\omega _{0}}{2}-\frac{U}{2N}\psi ^{\ast }\psi%
\end{array}%
\right) .  \label{MA}
\end{equation}

\begin{figure}[t]
\includegraphics[width=8cm]{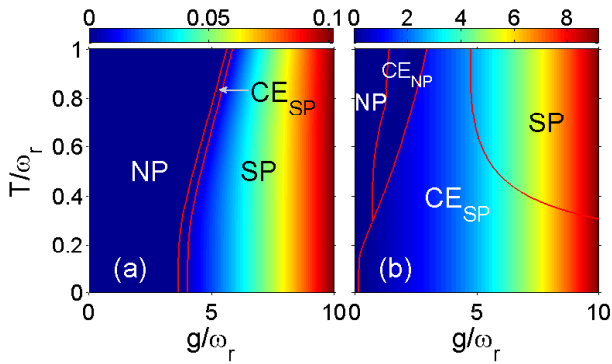}\newline
\caption{(Color online) The scaled mean-photon number as the functions of
the temperature $T$ and the collective coupling strength $g$ for $U=10.00$ $%
\protect\omega _{r}$, $\Delta _{c}=-20.00\protect\omega _{r}$ (a) and $%
U=31.10$ $\protect\omega _{r}$, $\Delta _{c}=20.00\protect\omega _{r}$ with $%
\protect\omega _{r}=2\protect\pi \times 3.71$ kHz.}
\label{fig3}
\end{figure}

We first integrate out the Fermi field $\Phi _{i}\left( \tau \right) $ (thus
the atom spin degrees of freedom) in Eq. (\ref{PT}), and then use the
standard stationary phase approximation \cite{Alexander}, $i.e.$, $\delta
S\left( \psi ^{\ast },\psi \right) /\delta \psi \left( \tau \right) =0$ and $%
\delta S\left( \psi ^{\ast },\psi \right) /\delta \psi ^{\ast }\left( \tau
\right) =0$, to obtain the finite temperature equilibrium phase diagram in
the large atom number, which is governed by the following equation
\begin{equation}
\omega \psi =F(\psi ^{\ast },\psi )\tanh \left[ \beta G(\psi ^{\ast },\psi )%
\right] \psi ,  \label{CTE}
\end{equation}%
and its complex conjugate. Here $F(\psi ^{\ast },\psi )=\left[ \left( \frac{%
\omega _{0}}{2}+\frac{U}{2N}\psi ^{\ast }\psi \right) U+4g^{2}\right] /[2%
\sqrt{\left( \frac{\omega _{0}}{2}+\frac{U}{2N}\psi ^{\ast }\psi \right)
^{2}+4\frac{g^{2}}{N}\psi ^{\ast }\psi }]$ and $G(\psi ^{\ast },\psi )=\sqrt{%
\left( \frac{\omega _{0}}{2}+\frac{U}{2N}\psi ^{\ast }\psi \right) ^{2}+4%
\frac{g^{2}}{N}\psi ^{\ast }\psi }/2$. It is straightforward to find that
Eq. (\ref{CTE}) has a trivial solution $\psi =\psi ^{\ast }=0$ and the
non-trivial solution $\psi \left( \tau \right) =$ $\psi ^{\ast }\left( \tau
\right) =\pm \psi _{0}$, satisfying a nonlinear equation
\begin{equation}
\frac{2\omega \zeta }{\left( \omega _{0}+U\bar{\psi}_{0}^{2}\right) U+8g^{2}}%
=\tanh \left( \frac{\beta \zeta }{4}\right) ,  \label{CES}
\end{equation}%
where $\zeta =\sqrt{\left( \omega _{0}+U\bar{\psi}_{0}^{2}\right)
^{2}+16g^{2}\bar{\psi}_{0}^{2}}$, and $\bar{\psi}_{0}^{2}=\left\langle \psi
^{\dagger }\psi \right\rangle /N$ is the scaled mean photon number. If $\psi
=\psi ^{\ast }=0$, the system is in the NP with no collective excitation.
However, the existence of non-trivial solutions shows that the macroscopic
collective excitation for both atoms and photon occurs and thus the system
locates at the SP. Finally, the stable conditions $\delta ^{2}S\left( \psi
^{\ast },\psi \right) /\delta \psi ^{2}>0$ and $\delta ^{2}S\left( \psi
^{\ast },\psi \right) /\delta (\psi ^{\ast })^{2}>0$ determine the real
solution of Eq. (\ref{CTE}) and thus the phases of the BEC-cavity system at
finite temperature.

When $U=0$, Eq. (\ref{CES}) reduces to $\omega \sqrt{\omega _{0}^{2}+16g^{2}%
\bar{\psi}_{0}^{2}}/4g^{2}=\tanh \left( \beta \sqrt{\omega _{0}^{2}+16g^{2}%
\bar{\psi}_{0}^{2}}/4\right) $, the result for the original Dicke model. In
this case, only SP and NP exist, even at finite temperature \cite{Popov}.
When the temperature $T\rightarrow 0$, $\tanh \left( \beta \zeta /4\right)
=1 $, and Eq. (\ref{CES}) becomes $2\omega \zeta /[\left( \omega _{0}+U\bar{%
\psi}_{0}^{2}\right) U+8g^{2}]=1$, from which rich zero-temperature phase
diagrams have been obtained \cite{Keeling,Bhaseen,Chen2}. With increasing
temperature $T$, the function $\tanh \left( \beta \zeta /4\right) $
decreases. As a consequence, Eq. (\ref{CES}) can possess some new physical
solutions of $\bar{\psi}_{0}$, leading to new phases at finite temperature.

\begin{figure}[t]
\includegraphics[width=8cm]{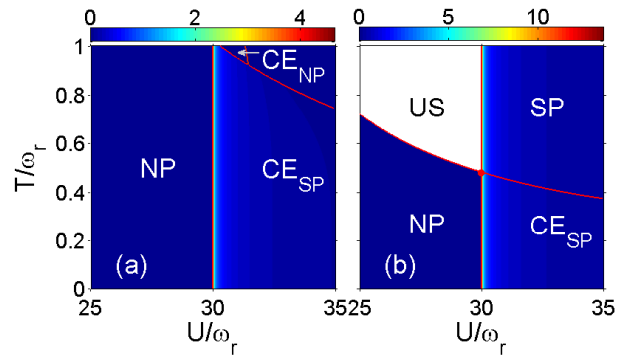}\newline
\caption{(Color online) The scaled mean-photon number as the functions of
the temperature $T$ and the nonlinear interaction $U$ for $g=3.00$ $\protect%
\omega _{r}$, $\Delta _{c}=20.00\protect\omega _{r}$ (a) and $g=6.00$ $%
\protect\omega _{r}$, $\Delta _{c}=20.00\protect\omega _{r}$ with $\protect%
\omega _{r}=2\protect\pi \times 3.71$ kHz.}
\label{fig4}
\end{figure}

We solve the nonlinear equation (\ref{CTE}), together with stable conditions
$\delta ^{2}S\left( \psi ^{\ast },\psi \right) /\delta \psi ^{2}>0$ and $%
\delta ^{2}S\left( \psi ^{\ast },\psi \right) /\delta (\psi ^{\ast })^{2}>0$%
, for various different physical parameters and calculate the scaled mean
photon number $\left\langle \psi ^{\dagger }\psi \right\rangle /N=\bar{\psi}%
_{0}^{2}$. In Fig. 2, we plot the scaled mean-photon number $\bar{\psi}%
_{0}^{2}$ as the functions of the detuning $\Delta _{c}$ and the square of
the collective coupling strength $g$ for different temperatures (a) $T=0.00$
nK, (b) $T=0.13$ nK, (c) $T=50.00$ nK, and (d) $T=150.00$ nK. To compare
with the experiment, we also choose the other parameters to be the same as
those in Ref. \cite{Baumann1}, that is, the nonlinear interaction $%
U=1.70\times 10^{3}\omega _{r}$, and the recoil frequency $\omega _{r}=2\pi
\times 3.71$ kHz. When $T=0$, due to the existence of the strong nonlinear
interaction, the CE$_{\text{NP}}$ phase is found, apart from the NP and SP,
as shown in Fig. 2 (a). In the CE$_{\text{NP}}$ phase, $\bar{\psi}%
_{0}^{2}\sim 0$, and almost no photon signature is generated. However, at
the experimental operating temperature $T=50.00$ nk, numerical solution of
Eq. (\ref{CTE}) shows that the finite-temperature critical point that
separates the NP and SP agrees with that in the experiment. More
interestingly, the CE$_{\text{NP}}$ phase becomes the CE$_{\text{SP}}$
phase, in which a weak photon signature (compared with the strong
superradiant regime) emerges. We note that such weak photon signature for
the CE$_{\text{SP}}$ phase has already be observed in the experiment in the
same parameter region, although it has not been explicitly pointed out. The
transition temperature from the CE$_{\text{NP}}$ phase at zero temperature
to the CE$_{\text{SP}}$ at $50$ nk is $T_{p}=0.13$ nk, as shown in Fig.
2(b). With increasing temperature $T$ (=$150.00$ nk), the SP region becomes
narrower due to the suppression of the superradiance by thermal
fluctuations, as expected. However, a new phase diagram including the NP,
SP, CE$_{\text{SP}}$, CE$_{\text{NP}}$ and US ($\delta ^{2}S\left( \psi
^{\ast },\psi \right) /\delta \psi ^{2}<0$ and $\delta ^{2}S\left( \psi
^{\ast },\psi \right) /\delta (\psi ^{\ast })^{2}<0$) is observed, as shown
in Fig. 2(d).

\begin{figure}[t]
\includegraphics[width=6cm]{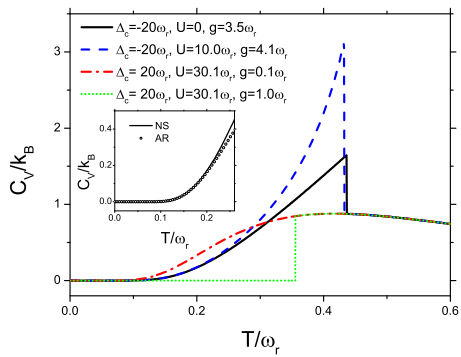}\newline
\caption{(Color online) The specific heat $C_{\text{v}}$ as a function of
the temperature $T$ for different nonlinear interaction $U$, detuning $%
\Delta _{c}$ and collective coupling strength $g$. Insert: comparsion
between the numerical simulation (NS) and the analytical result (AR) in Eq. (%
\protect\ref{HCSP}) for the specific heat $C_{\text{v}}$. }
\label{fig5}
\end{figure}

In a realistic experiment, the experimental parameters such as the detuning $%
\Delta _{c}$, the collective coupling strength $g$, the nonlinear
interaction $U$ as well as the temperature $T$ can be controlled
independently. For example, $\Delta _{c}$ and $g$ can be tuned by adjusting
the frequency and power of the pump laser, respectively. The nonlinear
interaction can be controlled by tuning both the frequency of the pump laser
and the atom number. The temperature can be manipulated by controlling the
evaporative cooling process for the BEC \cite{REV}. In Fig. 3, we plot the
scaled mean-photon number $\bar{\psi}_{0}^{2}$ as a function of the
temperature $T$ and the collective coupling strength $g$. While in Fig. 4,
we plot the scaled mean-photon number $\bar{\psi}_{0}^{2}$ as a function of
the temperature $T$ and the nonlinear interaction $U$. These figures show
that rich finite-temperature phase diagrams exist in different parameter
regions, induced by the competition among these parameters. Interestingly,
we find a four-phase coexistence point in certain parameter region, as shown
in Fig. \ref{fig4}(b). It should be emphasized again that these exotic
phases arise from the interesting nonlinear interaction $U$ and the finite
temperature. If $U=0$, only the SP and NP can be found \cite{Popov}.

These rich finite temperature phases and the phase transitions can be
detected in experiments by measuring not only the mean photon number, but
also some the thermodynamic quantities, such as the specific heat per atom $%
C_{\text{V}}=\frac{1}{Nk_{B}T^{2}}\frac{\partial ^{2}}{\partial \beta ^{2}}%
\left( \ln Z\right) $ in the BEC. In the NP, $C_{\text{V}}$ can be obtained
analytically, yielding $C_{\text{V}}^{\text{NP}}=\omega _{0}^{2}\left[
1-\tanh ^{2}\left( \beta \omega _{0}/4\right) \right] /(8k_{B}T^{2})$ \cite%
{Hepp}, which reaches the maximum at certain temperature (the red
dashed-dotted line in Fig. 5). However, the explicit expression for $C_{%
\text{V}}$ cannot be obtained in the SP. Nevertheless, in the region $T\ll
T_{c}$, we find
\begin{equation}
C_{\text{V}}^{\text{SP}}\simeq \frac{\zeta _{0}^{2}}{8k_{B}T^{2}}\text{sech}%
^{2}(\frac{\zeta _{0}}{4k_{B}T})  \label{HCSP}
\end{equation}%
where $\zeta _{0}$ is the value of $\zeta =\sqrt{\left( \omega _{0}+U\bar{%
\psi}_{0}^{2}\right) ^{2}+16g^{2}\bar{\psi}_{0}^{2}}$ at $T=0$. This
expression shows that the specific heat $C_{\text{V}}^{\text{SP}}$ in the SP
increases rapidly (exponential increase at low temperature), which agrees
well with the numerical results (see the insert of Fig. 5). At the critical
temperature $T_{c}$, at which the system enters the NP, the specific heat $%
C_{\text{V}}$ has a large jump. This step behavior is quite different from
the behavior of the scaled mean photon number $\bar{\psi}_{0}^{2}$,\textbf{\
}which varies smoothly when crossing the critical point. Therefore the
temperature-driven phase transition from SP to NP can be detected using the
specific heat $C_{\text{V}}$. In addition, the behavior of specific heat $C_{%
\text{V}}$ in the CE$_{\text{SP}}$/CE$_{\text{NP}}$ is similar to that in
the SP/NP. Moreover, for the transition from the CE$_{\text{SP}}$\ to CE$_{%
\text{NP}}$\ phases, such a large jump of $C_{\text{V}}$\ still exists, as
shown by the green dotted line of Fig. 5.

In summary, we develop a finite temperature theory for the Dicke phase
transition of a BEC in an optical cavity. Our theory not only explains well
the recent experimental observation of the phase transition from the SP to
NP, but predicts rich new phases that either have been observed (but not
pointed out) or may be observable by tuning experimental parameters. Our
study is crucially important for understanding the Dicke phase transition in
a realistic experimental system and may have significant application in
quantum computation, quantum optics, etc.

We thank Profs. T. Esslinger and Jing Zhang, and Drs. K. Baumann, F.
Brennecke, Yongping Zhang and Ming Gong for helpful discussions and
suggestions. This work is supported by the 973 Program under Grant
No. 2012CB921603, the NNSFC under Grant Nos. 10934004, 60978018,
61008012, 11074154, and 11075099. C.Z. is supported by NSF.

\end{document}